\documentclass[twocolumn,showpacs,showkeys,preprintnumbers]{revtex4}
\usepackage{graphicx}
\usepackage{dcolumn}
\usepackage{bm}

\begin{document}

\title{Born-Infeld gravity in any dimension}
\author{J. A. Nieto}
\affiliation{Facultad de Ciencias F\'{\i}sico-Matem\'aticas de la Universidad Aut\'onoma
de Sinaloa, Av. de las Americas y Blvd. Universitarios, Ciudad
Universitaria, C.P. 80010, Culiac\'{a}n Sinaloa, M\'{e}xico}
\email{nieto@uas.uasnet.mx}
\date{June 17 2004}

\begin{abstract}
We develop a Born-Infeld type theory for gravity in any dimension. We show
that in four dimensions our formalism allows a self-dual (or anti-self dual)
Born-Infeld gravity description. Moreover, we show that such a self-dual
action is reduced to both the Deser-Gibbons and the Jacobson-Smolin-Samuel
action of Ashtekar formulation. A supersymmetric generalization of our
approach is outlined.
\end{abstract}

\maketitle







\section{INTRODUCTION}

More than 80 years ago, Eddington [1] introduced the action

\begin{equation}
S_{EDD}=\int d^{4}x\sqrt{\det (R_{\mu \nu }(\Gamma ))},  \label{1}
\end{equation}%
as a possibility to incorporate affine invariance in a gravitational
context. Here $R_{\mu \nu }=R_{\nu \mu }$ is the Ricci tensor and $\Gamma
_{\mu \nu }^{\alpha }=\Gamma _{\nu \mu }^{\alpha }$ becomes the Christoffel
symbol, after using the Palatini method.

In 1934 Born and Infeld (BI) [2] for different reasons than Eddington
proposed a similar action for electrodynamics

\begin{equation}
S_{BI}=\int d^{4}x\sqrt{\det (g_{\mu \nu }+F_{\mu \nu })},  \label{2}
\end{equation}%
where $g_{\mu \nu }$ is the space-time metric and $F_{\mu \nu }$ is the
electromagnetic field strength. It turns out that the action $S_{BI}$
reduces to Maxwell action for small amplitudes. Some years ago, the action
(2) became of very much interest because its relation with SUSY (see [3] and
references there in) and string theory (see [4]-[7] and references there in)
and D-brane physics [8]. At present, the BI theory has been generalized to
the supersymmetric non abelian case (see [9] and references there in) and it
has been connected with noncommutative geometry [10] and Cayley-Dickson
algebras [11]. Moreover, it turns out intriguing that the BI action can be
derived from matrix theory [12] and D9-brane [13]. For interesting comments
and observation about the BI theory the reader is refer to the works of
Schwarz [14] and Ketov [15].

Inspired by the interesting properties of the BI theory, Deser and Gibbons
[16] proposed in 1998 the analogue of $S_{BI}$ for a gravity,

\begin{equation}
S_{DG}=\int d^{4}x\sqrt{\det (ag_{\mu \nu }+bR_{\mu \nu }+cX_{\mu \nu })}.
\label{3}
\end{equation}%
Here, $X_{\mu \nu }=X_{\nu \mu }$ stands for higher order corrections in
curvature and $a,b$ and $c$ are appropriate coupling constants. Deser and
Gibbons determined some possible choices for $X_{\mu \nu }$ by imposing
three basic criteria to the action (3), namely freedom of ghosts,
regularization of singularities and supersymmetrizability. In fact, they
found that the expression $X_{\mu \nu }\sim R_{\mu }^{\alpha }R_{\alpha \nu }
$ may provide one interesting choice for $X_{\mu \nu },$ allowing (3) to
describe gravitons but no ghosts.

From the point of view of string theory both the photon as well as the
graviton should be part of the spectrum of a string. Therefore one should
expect that just as $S_{BI}$ is used to regularize $p$-branes the $S_{DG}$
action may be used with similar purpose. However, for this possibility to be
viable it is necessary to generalize the $S_{DG}$ action to higher
dimensions. Moreover, Wohlfarth [17] has pointed out that the
supersymmetrizability requirement presumably implied by M-theory, may
provide another reason to be interested in such a higher-dimensional
extension of $S_{DG}.$

In this work, we use similar technics as the one used in MacDowell-Mansouri
formalism [18] (see also [19] and references there in) in order to
generalize the $S_{DG}$ action to higher dimensions. In four dimensions, our
approach is reduced to the theory of Deser-Gibbons. We also show that in
four dimensions our action admits self-dual (antiself-dual) generalization
which is reduced to the Ashtekar action as proposed by
Jacobson-Smolin-Samuel [20]. Since Born-Infeld type action has shown to be
also useful to study different aspects of the brane world scenario, our work
may appear interesting in this area of research.

The plan of this paper is as follows: In section 2, using a
MacDowell-Mansouri's type formalism we establish our proposed action for
Born-Infeld gravity in any dimension. In section 3, we discuss the
relationship between our proposed action and the Deser and Gibbons action.
In section 4, we consider a self-dual (antiself-dual) version of our
proposed action in four dimensions and we show that it is reduced to the
Jacobson-Smolin-Samuel action for the Ashtekar variables. Finally, in
section 5 we outline a possible supersymmetric generalization of our
formalism and we make some final remarks.

\bigskip\

\noindent \textbf{2.- PROPOSED BORN-INFELD-GRAVITY (BIG) ACTION}

\smallskip\

Consider the extended curvature
\begin{equation}
\mathcal{R}_{\mu \nu }^{ab}=R_{\mu \nu }^{ab}+\Sigma _{\mu \nu }^{ab},
\label{4}
\end{equation}%
where%
\begin{equation}
R_{\mu \nu }^{ab}=\partial _{\mu }\omega _{\nu }^{ab}-\partial _{\nu }\omega
_{\mu }^{ab}+\omega _{\mu }^{ac}\omega _{\nu c}^{\quad b}-\omega _{\nu
}^{ac}\omega _{\mu c}^{\quad b}  \label{5}
\end{equation}%
and%
\begin{equation}
\Sigma _{\mu \nu }^{ab}=-(e_{\mu }^{a}e_{\nu }^{b}-e_{\nu }^{a}e_{\mu }^{b}).
\label{6}
\end{equation}%
Here, $\omega _{\mu }^{ab}$ is a $SO(d-1,1)$ connection and $e_{\mu }^{a}$
is a vierbein field. It turns out that (4) can be obtained using the
MacDowell-Mansouri formalism [18] (see also Ref. [19] and references there
in). In fact, in such a formalism the gravitational field is represented as
a connection one-form associated to some group which contains the Lorentz
group as a subgroup. The typical example is provided by the $SO(d,1)$ de
Sitter gauge theory of gravity. In this case, the $SO(d,1)$ gravitational
gauge field $\omega _{\mu }^{AB}=-$ $\omega _{\mu }^{BA}$ is broken into the
$SO(d-1,1)$ connection $\omega _{\mu }^{ab}$ and the $\omega _{\mu
}^{da}=e_{\mu }^{a}$ vierbein field, with the dimension $d$ fixed. Thus, the
de Sitter (or anti-de Sitter) curvature
\begin{equation}
\mathcal{R}_{\mu \nu }^{AB}=\partial _{\mu }\omega _{\nu }^{AB}-\partial
_{\nu }\omega _{\mu }^{AB}+\omega _{\mu }^{AC}\omega _{\nu C}^{\quad
B}-\omega _{\nu }^{AC}\omega _{\mu C}^{\quad B}  \label{7}
\end{equation}%
leads to the curvature (4).

Let us now introduce the definition

\begin{equation}
\mathcal{R}_{\mu }^{a}\equiv e_{b}^{\nu }\mathcal{R}_{\mu \nu }^{ab},
\label{8}
\end{equation}%
where $e_{b}^{\nu }$ is the inverse vierbien field.

Our proposed action is%
\begin{equation}
S=-\frac{1}{d!}\int d^{d}x{}\varepsilon ^{\mu _{1}...\mu _{d}}\epsilon
_{a_{1}...a_{d}}\mathcal{R}_{\mu _{1}}^{a_{1}}{}...\mathcal{R}_{\mu
_{d}}^{a_{d}},  \label{9}
\end{equation}%
where $\varepsilon ^{\mu _{1}...\mu _{d}}$ is the completely antisymmetric
tensor associated to the space-time, with $\varepsilon ^{0...d-1}=1$ and $%
\varepsilon _{0...d-1}=1$, while $\epsilon _{a_{1}...a_{d}}$ is also the
completely antisymmetric tensor but now associated to the internal group $%
S(d-1,1)$, with $\epsilon _{0...d-1}=-1.$ We assume that the internal metric
is given by $(\eta _{ab})=diag(-1,...,1)$. So, we have $\epsilon
^{0...d-1}=1 $.

Using (6) we get%
\begin{equation}
e_{b}^{\nu }\Sigma _{\mu \nu }^{ab}=\lambda e_{\mu }^{a},  \label{10}
\end{equation}%
where $\lambda =1-d.$ Therefore, (4) and (8) and (10) lead to

\begin{equation}
\mathcal{R}_{\mu }^{a}=R_{\mu }^{a}+\lambda e_{\mu }^{a},  \label{11}
\end{equation}%
where

\begin{equation}
R_{\mu }^{a}\equiv e_{b}^{\nu }R_{\mu \nu }^{ab}.  \label{12}
\end{equation}

Substituting (11) into (9) we obtain

\begin{equation}
S=-\frac{1}{d!}\int d^{d}x{}\varepsilon ^{\mu _{1}...\mu _{d}}\epsilon
_{a_{1}...a_{d}}(R_{\mu _{1}}^{a_{1}}+\lambda e_{\mu
_{1}}^{a_{1}})...(R_{\mu _{d}}^{a_{d}}+\lambda e_{\mu _{d}}^{a_{d}}).
\label{13}
\end{equation}%
From this action we get

\begin{equation}
\begin{array}{c}
S=-\frac{\lambda ^{d}}{d!}\int d^{d}x\varepsilon ^{\mu _{1}...\mu
_{d}}\epsilon _{a_{1}...a_{d}}e_{\mu _{1}}^{a_{1}}...e_{\mu _{d}}^{a_{d}} \\
\\
-\frac{d\lambda ^{d-1}}{d!}\int d^{d}x\varepsilon ^{\mu _{1}...\mu
_{d}}\epsilon _{a_{1}...a_{d}}e_{\mu _{1}}^{a_{1}}...e_{\mu
_{d-1}}^{a_{d-1}}R_{\mu _{d}}^{a_{d}} \\
\\
-\frac{d(d-1)\lambda ^{d-2}}{2!d!}\int d^{d}x\varepsilon ^{\mu _{1}...\mu
_{d}}\epsilon _{a_{1}...a_{d}}e_{\mu _{1}}^{a_{1}}...e_{\mu
_{d-2}}^{a_{d-2}}R_{\mu _{d-1}}^{a_{d-1}}R_{\mu _{d}}^{a_{d}} \\
\\
-...-\frac{1}{d!}\int d^{d}x\varepsilon ^{\mu _{1}...\mu _{d}}\epsilon
_{a_{1}...a_{d}}R_{\mu _{1}}^{a_{1}}...R_{\mu _{d}}^{a_{d}}.%
\end{array}
\label{14}
\end{equation}%
Since

\begin{equation}
\varepsilon ^{\mu _{1}...\mu _{d}}\epsilon
_{a_{1}...a_{d}}=-e(e_{a_{1}}^{[\mu _{1}}...e_{a_{d}}^{\mu _{d}]}),
\label{15}
\end{equation}%
where the bracket $[\mu _{1}...\mu _{d}]$ means completely antisymmetric and
$e=\det (e_{\mu }^{a})$, we find that (14) can be written as

\begin{equation}
\begin{array}{c}
S=\lambda ^{d}\int d^{d}xe+\lambda ^{d-1}\int d^{d}xeR \\
\\
+\frac{\lambda ^{d-2}}{2!}\int d^{d}xe(R^{2}-R^{\mu \nu }R_{\mu \nu
})+...+\int d^{d}x{}\det (R_{\mu }^{a}).%
\end{array}
\label{16}
\end{equation}%
Here, $R\equiv e_{a}^{\mu }R_{\mu }^{a}$ and $R_{\mu \nu }\equiv e_{\nu
}^{a}R_{\mu a},$while $R^{\mu \nu }\equiv g^{\mu \alpha }g^{\nu \beta
}R_{\alpha \beta },$ where $g^{\alpha \beta }$ is the matrix inverse of $%
g_{\alpha \beta }\equiv e_{\alpha }^{a}e_{\beta }^{b}\eta _{\alpha \beta }.$
We recognize in the first and second terms in (16) the Einstein-Hilbert
action with cosmological constant in $d$-dimensions.

\bigskip\

\noindent \textbf{3.- RELATION WITH \ DESER-GIBBONS TYPE ACTION}

\smallskip\

Let us define the tensor

\begin{equation}
G_{\mu \nu }\equiv \frac{1}{\lambda ^{2}}\mathcal{R}_{\mu }^{a}\mathcal{R}%
_{\nu }^{b}\eta _{ab}.  \label{17}
\end{equation}%
Using (11) we get

\begin{equation}
\begin{array}{c}
G_{\mu \nu }=\frac{1}{\lambda ^{2}}(R_{\mu }^{a}+\lambda e_{\mu
}^{a})(R_{\nu }^{b}+\lambda e_{\nu }^{b})\eta _{ab} \\
\\
=g_{\mu \nu }+\Lambda R_{\mu \nu }+\frac{\Lambda ^{2}}{4}R_{\mu }^{\alpha
}R_{\alpha \nu },%
\end{array}
\label{18}
\end{equation}%
where $\Lambda =\frac{2}{\lambda }$. Here, we used the fact that $R_{\mu \nu
}=R_{\nu \mu }$.

From (17), it is not difficult to see that

\begin{equation}
\det (G_{\mu \nu })\equiv -\frac{1}{\lambda ^{2d}}\mathcal{R}^{2},
\label{19}
\end{equation}%
where $\mathcal{R=}\det (\mathcal{R}_{\mu }^{a})$. Therefore, our proposed
action (9) can be expressed in terms of $G_{\mu \nu }$ as

\begin{equation}
\begin{array}{c}
S=-\frac{1}{d!}\int d^{d}x\varepsilon ^{\mu _{1}...\mu _{d}}{}\epsilon
_{a_{1}...a_{d}}\mathcal{R}_{\mu _{1}}^{a_{1}}...\mathcal{R}_{\mu
_{d}}^{a_{d}} \\
\\
=\int d^{d}x\mathcal{R}=\lambda ^{d}\int d^{d}x\sqrt{-\det (G_{\mu \nu })}.%
\end{array}
\label{20}
\end{equation}

In virtue of (18) the action (20) becomes

\begin{equation}
S=\frac{2^{d}}{\Lambda ^{d}}\int d^{d}x\sqrt{-\det (g_{\mu \nu }+\Lambda
R_{\mu \nu }+\frac{\Lambda ^{2}}{4}R_{\mu }^{\alpha }R_{\alpha \nu })},
\label{21}
\end{equation}%
which is a Born-Infeld type action for gravity in any dimension. In four
dimensions, we recognize that the action (21) corresponds to the one
proposed by Deser and Gibbons [16] (see Eq. (3) of [16]), with $X_{\mu \nu
}=R_{\mu }^{\alpha }R_{\alpha \nu }$.

\bigskip\

\noindent \textbf{4.- SELF-DUAL (ANTISELF-DUAL) ACTION IN FOUR DIMENSIONS}%
\smallskip\

\smallskip\

In four dimensions the action (9) can be generalized to the case of
self-dual (antiself-dual) gauge gravitational field. We define the self-dual
(antiself-dual) of $\mathcal{R}_{\mu \nu }^{cd}$ as

\begin{equation}
^{\pm }\mathcal{R}_{\mu \nu }^{ab}=\frac{1}{2}^{\pm }M_{cd}^{ab}\mathcal{R}%
_{\mu \nu }^{cd},  \label{22}
\end{equation}%
where

\begin{equation}
^{\pm }M_{cd}^{ab}=\frac{1}{2}(\delta _{cd}^{ab}\mp i\epsilon _{cd}^{ab}).
\label{23}
\end{equation}%
Here, $\delta _{cd}^{ab}=\delta _{c}^{a}\delta _{d}^{b}-\delta
_{c}^{b}\delta _{d}^{a}$. We also define

\begin{equation}
^{\pm }\mathcal{R}_{\mu }^{a}\equiv e_{b}^{\nu \pm }\mathcal{R}_{\mu \nu
}^{ab}.  \label{24}
\end{equation}

In four dimensions (9) becomes

\begin{equation}
S=-\frac{1}{4!}\int d^{4}x{}\varepsilon ^{\mu _{1}...\mu _{4}}{}\epsilon
_{a_{1}...a_{4}}\text{ }\mathcal{R}_{\mu _{1}}^{a_{1}}{}...\mathcal{R}_{\mu
_{4}}^{a_{4}}.  \label{25}
\end{equation}%
Substituting (11) into this expression lead us to

\begin{equation}
\begin{array}{c}
S=-\frac{1}{4!}\int d^{4}x\varepsilon ^{\mu _{1}...\mu _{4}}\epsilon
_{a_{1}...a_{4}}(R_{\mu _{1}}^{a_{1}}+\lambda e_{\mu
_{1}}^{a_{1}})...(R_{\mu _{4}}^{a_{4}}+\lambda e_{\mu _{4}}^{a_{4}}) \\
\\
=-\frac{\lambda ^{4}}{4!}\int d^{4}x\varepsilon ^{\mu _{1}...\mu
_{4}}\epsilon _{a_{1}...a_{4}}e_{\mu _{1}}^{a_{1}}...e_{\mu _{4}}^{a_{4}} \\
\\
-\frac{4\lambda ^{3}}{4!}\int d^{4}x\varepsilon ^{\mu _{1}\mu _{2}\mu
_{3}\mu _{4}}\epsilon _{a_{1}a_{2}a_{3}a_{4}}e_{\mu _{1}}^{a_{1}}e_{\mu
_{2}}^{a_{2}}e_{\mu _{3}}^{a_{3}}R_{\mu _{4}}^{a_{4}} \\
\\
-\frac{6\lambda ^{2}}{4!}\int d^{4}x\varepsilon ^{\mu _{1}\mu _{2}\mu
_{3}\mu _{4}}\epsilon _{a_{1}a_{2}a_{3}a_{4}}e_{\mu _{1}}^{a_{1}}e_{\mu
_{2}}^{a_{2}}R_{\mu _{3}}^{a_{3}}R_{\mu _{4}}^{a_{4}} \\
\\
-\frac{4\lambda }{4!}\int d^{4}x\varepsilon ^{\mu _{1}\mu _{2}\mu _{3}\mu
_{4}}\epsilon _{a_{1}a_{2}a_{3}a_{4}}e_{\mu _{1}}^{a_{1}}R_{\mu
_{2}}^{a_{2}}R_{\mu _{3}}^{a_{3}}R_{\mu _{4}}^{a_{4}} \\
\\
-\frac{1}{4!}\int d^{4}x\varepsilon ^{\mu _{1}\mu _{2}\mu _{3}\mu
_{4}}\epsilon _{a_{1}a_{2}a_{3}a_{4}}R_{\mu _{1}}^{a_{1}}R_{\mu
_{2}}^{a_{2}}R_{\mu _{3}}^{a_{3}}R_{\mu _{4}}^{a_{4}}.%
\end{array}
\label{26}
\end{equation}%
Simplifying these expressions one finds

\begin{equation}
\begin{array}{c}
S=\lambda ^{4}\int d^{4}xe+\lambda ^{3}\int d^{4}xeR \\
\\
+\frac{\lambda ^{2}}{2!}\int d^{4}xe(R^{2}-R^{\mu \nu }R_{\mu \nu }) \\
\\
+\frac{\lambda }{3!}\int d^{4}xe(R^{3}-3RR^{\mu \nu }R_{\mu \nu }+2R^{\mu
\alpha }R_{\alpha \beta }R_{\mu }^{\beta }) \\
\\
+\int d^{4}xe\det (R_{\mu \nu }).%
\end{array}
\label{27}
\end{equation}

Let us now generalize (25) in the form

\begin{equation}
S=-\frac{1}{4!}\int d^{4}x\varepsilon ^{\mu _{1}...\mu _{4}}\epsilon
_{a_{1}...a_{4}}{}^{\pm }\mathcal{R}_{\mu _{1}}^{a_{1}}...^{\pm }\mathcal{R}%
_{\mu _{4}}^{a_{4}}.  \label{28}
\end{equation}%
It is not difficult to see that

\begin{equation}
^{\pm }\mathcal{R}_{\mu }^{a}\equiv ^{\pm }R_{\mu }^{a}+\lambda e_{\mu }^{a}.
\label{29}
\end{equation}%
Thus, we have

\begin{equation}
S=-\frac{1}{4!}\int d^{4}x\varepsilon ^{\mu _{1}...\mu _{4}}\epsilon
_{a_{1}...a_{4}}(^{\pm }R_{\mu _{1}}^{a_{1}}+\lambda e_{\mu
_{1}}^{a_{1}})...(^{\pm }R_{\mu _{4}}^{a_{4}}+\lambda e_{\mu _{4}}^{a_{4}}).
\label{30}
\end{equation}%
Therefore, we obtain

\begin{equation}
\begin{array}{c}
S=\lambda ^{4}\int d^{4}xe+\lambda ^{3}\int d^{4}xe^{\pm }R \\
\\
+\frac{\lambda ^{2}}{2!}\int d^{4}xe(^{\pm }R^{2}-^{\pm }R^{\mu \nu \pm
}R_{\mu \nu }) \\
\\
+\frac{\lambda }{3!}\int d^{4}xe(^{\pm }R^{3}-3^{\pm }R^{\pm }R^{\mu \nu \pm
}R_{\mu \nu }+2^{\pm }R^{\mu \alpha \pm }R_{\alpha \beta }{}^{\pm }R_{\mu
}^{\beta }) \\
\\
+\int d^{4}xe\det (^{\pm }R_{\mu \nu }).%
\end{array}
\label{31}
\end{equation}%
We recognize in the second term of this expression the Ashtekar action as
proposed by Jacobson-Smolin-Samuel [20].

\bigskip\

\bigskip\

\noindent \textbf{5.-FINAL COMMENTS}

\smallskip\

In this work, we proposed the action (9) as an alternative to generalize to
higher dimensions the Born-Infeld-gravity action of Deser and Gibbons. We
proved that our action in four dimensions allows a generalization to
self-dual (antiself-dual) action. Moreover, we showed that such a self-dual
action is reduced to the Jacobson-Smolin-Samuel action of the Ashtekar
formulation.

It is remarkable that the only requirement for the proposed action (9) is
the general covariance implied by the transition $SO(d,1)\rightarrow
SO(d-1,1).$ It remains to check whether the action (9) satisfies the
criteria of ghosts freedom and regularization of singularities used by Deser
and Gibbons [16] in four dimensions and Wohlfarth [17] in higher dimensions.
However, these requirements should be applied very carefully to (9) because
under compactification (9) should lead not only to gravity but also to
Yang-Mills and scalar fields. From the point of view of M-theory one may
even think in a phase in which matter fields and ghosts are mixed implying a
generalization of (9) to some kind of topological theory.

As we mentioned in the introduction, supersymmetrizability provides to
Wohlfarth [17] with one of the main motivation for considering BI gravity in
higher dimensions. Wohlfarth's theory is constructed by using the symmetries
of the Riemann tensor, considering tensors with pair of antisymmetrized
indices. Specifically, the Wohlfarth's formulation is based on the action $%
\int \sqrt{-g}[(\det (\delta _{B}^{A}+\xi R_{B}^{A})^{\varsigma }-1],$ where
$R_{AB}=R_{[\mu \nu ][\alpha \beta ]},$ $\delta _{B}^{A}=\delta _{\mu
}^{\alpha }\delta _{\nu }^{\beta }-\delta _{\mu }^{\beta }\delta _{\nu
}^{\alpha }$ and $\xi $, $\varsigma $ are constants (see Ref. [17] for
details). Although interesting, it does not seem evident how to
supersymmetrize this theory.

In contrast, our treatment admits a straightforward generalization to the
supersymmetric version in four dimensions. In fact, in the action (9) we may
replace the curvature $\mathcal{R}_{\mu \nu }^{ab}$ by its supersymmetric
extension%
\begin{equation}
\begin{array}{l}
\mathcal{R}_{\mu \nu }^{ab}=R_{\mu \nu }^{ab}+\Sigma _{\mu \nu }^{ab}+\Theta
_{\mu \nu }^{ab}, \\
\\
\mathcal{R}_{\mu \nu }^{i}=R_{\mu \nu }^{i}+\Sigma _{\mu \nu }^{i}, \\
\\
\mathcal{R}_{\mu \nu }^{a}=R_{\mu \nu }^{a}+\Sigma _{\mu \nu }^{a},%
\end{array}
\label{32}
\end{equation}%
where

\begin{equation}
\Theta _{\mu \nu }^{ab}={\frac{1}{2}}f_{ij}^{ab}\psi _{\mu }^{i}\psi _{\nu
}^{j},  \label{33}
\end{equation}%
\begin{equation}
\Sigma _{\mu \nu }^{a}={\frac{1}{2}}f_{ij}^{4a}\psi _{\mu }^{i}\psi _{\nu
}^{j},  \label{34}
\end{equation}%
\begin{equation}
\Sigma _{\mu \nu }^{i}=f_{4aj}^{i}(e_{\mu }^{a}\psi _{\nu }^{j}-e_{\nu
}^{a}\psi _{\mu }^{j}),  \label{35}
\end{equation}%
and

\begin{equation}
R_{\mu \nu }^{i}=\partial _{\mu }\psi _{\nu }^{i}-\partial _{\nu }\psi _{\mu
}^{i}+\frac{1}{2}f_{cdj}^{i}(\omega _{\mu }^{cd}\psi _{\nu }^{j}-\omega
_{\nu }^{cd}\psi _{\mu }^{j}).  \label{36}
\end{equation}%
Here, $\psi _{\mu }^{i}$ is a spinor field and the quantities $f_{cdj}^{i}$
and so on are the structure constants of algebra associated to the Lie
supergroup $Osp(1\mid 4)$. It may be interesting for further research to see
if the resulting supersymmetric Born-Infeld gravity theory is related to the
$D=4$, $N=1$ Born-Infeld supergravity proposed by Gates and Ketov [21].

Since some authors [22]-[23] have found interesting connections between the
nonabelian Born-Infeld theory and $D$-branes it may also be interesting for
further research to investigate the relation of the action (9) with string
theory and with $p$-branes physics.

Recently, Liu and Li [24] have found an interesting application of
Born-Infeld and brane worlds and Garc\'{\i}a-Salcedo and Breton [25] have
studied the possibility that Born-Infeld inflates the Bianchi cosmological
models. It may be interesting to pursue an application of the action (9) in
these directions.

Pilatnik [26] has found spherical static solutions of the Born-Infeld
gravity theory which may also correspond to the theory described by the
action (9). Feigenbaum [27] found that Born regulates gravity in four
dimensions, so it appears attractive to see if the action (9) can be used to
regulate gravity in higher dimensions. Furthermore, Vollick [28] used the
Palatini formalism in Born-Infeld Einstein theory in four dimensions and
identify the antisymmetric part of the Ricci tensor with the electromagnetic
field. So, if we use similar Palatini's technics in connection with the
action (9) one should expect to identify part of the Ricci tensor with
generalized electromagnetic field, perhaps Yang-Mills field.

Finally, it has been shown (see [29]) that dualities in M-theory (see [30]
and references there in) are deeply related to Born Infeld theory. Since
eleven dimensional supergravity must be part of M-theory one should expect
that Born-Infeld gravity theory in any dimensions may play an important role
in this context. In this direction the derivations of Born-Infeld theory
from supergravity [31] and the Nambu-Goto action from Born-Infeld theory
[32] may be of particular interest to continue a further research in
connection with the action (9).

\bigskip\

\noindent \textbf{Acknowledgments: }I would like to thank to M. C. Mar\'{\i}%
n for helpful comments.


\bigskip

\begin{thebibliography}{99}
\bibitem{1} A. S. Eddington, \textit{The Mathematical Theory of Relativity}
(CUP 1924).

\bibitem{2} M. Born and L. Infeld, Proc. Roy. Soc. \textbf{A144}, 425 (1934).

\bibitem{3} A. Silvas "Born-Infeld and Supersymmetry", Ph.D. Thesis (la
Plata); hep-th/0012267.

\bibitem{4} A. A. Tseytlin, Nucl. Phys. \textbf{B501}, 41 (1997);
hep-th/9701125.

\bibitem{5} G. Karatheodoris, A. Pinzul and A. Stern, Mod. Phys. Lett.
\textbf{A18}, 1681 (2003).

\bibitem{6} I. Y. Park, Phys. Rev. \textbf{D64}, 081901 (2001);
hep-th/0106078.

\bibitem{7} G. W \ Gibbons, Rev. Mex. Fis. \textbf{49S1}, 19 (2003),
hep-th/0106059.

\bibitem{8} R. Emparan, Phys. Lett. \textbf{B423}, 71 (1998); hep-th/9711106.

\bibitem{9} E. A. Bergshoeff, A. Bilal, M. de Roo and A. Sevrin, JHEP
\textbf{0107}, 029 (2001); hep-th/0105274.

\bibitem{10} E. Serie, T. Masson and R. Kerner, "Nonabelian generalization
of Born-Infeld theory by noncommutative geometry";\ hep-th/0307105.

\bibitem{11} S. Kuwata, "Born-Infeld lagrangian using Cayley-Dickson
algebras"; hep-th/0306271.

\bibitem{12} E. Keski-Vakkuri and Per Kraus, Nucl. Phys. \textbf{B518}, 212
(1998); hep-th/9709122.

\bibitem{13} S. F. Kerstan, Class. Quant. Grav. \textbf{19}, 4525 (2002);
hep-th/0204225.

\bibitem{14} J. H. Schwarz, "Comments on Born-Infeld theory", talk given at
Strings 2001: International Conference, Mumbai, India, 5-10 Jan 2001;
hep-th/0103165.

\bibitem{15} S. V. Ketov, "Many faces of Born-Infeld theory", Invited talk
at 7th International Wigner Symposium (Wigsym 7), College Park, Maryland,
24-29 Aug 2001; hep-th/0108189.

\bibitem{16} S. Deser and G. W. Gibbons, Class. Quant. Grav. \textbf{15},
L35 (1998); hep-th/9803049.

\bibitem{17} M. N.R. Wohlfarth "Gravity a la Born-Infeld"; hep-th/0310067.

\bibitem{18} S. W. MacDowell and F. Mansouri, Phys. Rev. Lett. \textbf{38},
739 (1997); F. Mansouri, Phys. Rev. \textbf{D16}, 2456 (1977).

\bibitem{19} H. Grac\'{\i}a-Compe\'{a}n, J. A. Nieto, O. Obreg\'{o}n and C.
Ram\'{\i}rez, Phys. Rev. \textbf{D59}, 124003 (1999); J. A. Nieto, O. Obreg%
\'{o}n and J. Socorro, Phys. Rev. \textbf{D50}, R3583 (1994); J. A. Nieto,
J. Socorro and O. Obreg\'{o}n, Phys. Rev. Lett. \textbf{76}, 3482 (1996).

\bibitem{20} T. Jacobson and L. Smolin, Class. Quant. Grav. \textbf{5}, 583
(1988); J. Samuel, Pramana J. Phys. \textbf{28}, L429 (1987).

\bibitem{21} S. J. Jr. Gates and S. V. Ketov, Class. Quant. Grav. \textbf{18}%
, 3561 (2001); hep-th/0104223.

\bibitem{22} S. Nagoaka, "Fluctuation analysis of the nonAbelian Born-Infeld
action in the background intersecting D-branes"; hep-th/0307232.

\bibitem{23} S. Stieberger and T. R. Taylor, Nucl. Phys. \textbf{B648}, 3-34
(2003); hep-th/0209064 .

\bibitem{24} D. Liu and X. Li, Phys. Rev. \textbf{D68}, 067301 (2003);
hep-th/0307239.

\bibitem{25} R. Garcia-Salcedo and N. Breton, "Born-Infeld inflates Bianchi
cosmologies"; hep-th/0212130

\bibitem{26} D. Palatnik, Phys. Lett. \textbf{B432}, 287 (1998);
quant-ph/9701017; Phys. Lett. B432, 287 (1998); quant-ph/970101

\bibitem{27} J. A. Feigenbaum, Phys. Rev. \textbf{D58}, 124023 (1998);
hep-th/9807114.

\bibitem{28} Dan N. Vollick, "Palatini approach to Born-Infeld-Einstein
theory and a geometric description of electrodynamics"; gr-qc/0309101.

\bibitem{29} D. M. Brace, "Duality in M-theory and Born Infeld theory",
Ph.D.Thesis (Advisor: Bruno Zumino).

\bibitem{30} M. J. Duff, Int. J. Mod. Phys. \textbf{A\ 11,} 5623 (1996),
hep-th/9608117.

\bibitem{31} M. Sato and A. Tsuchiya, Prog. Theor. Phys. \textbf{109}, 687
(2003); hep-th/0211074

\bibitem{32} S. Ansoldi, C. Castro, E. I. Guendelman and E. Spallucci,
Class. Quant. Grav. \textbf{19}, L135 (2002); hep-th/0201018.
\end{thebibliography}
\end{document}